\documentclass[useAMS,usenatbib]{mn2e}

\usepackage{graphicx}
\usepackage{color}

\title[Mass ejection by PPI and luminous SNe]{Mass ejection by pulsational pair instability in very massive stars and implications for luminous supernovae}
\author[T. Yoshida, et al.]{Takashi Yoshida$^{1}$\thanks{E-mail:
tyoshida@astron.s.u-tokyo.ac.jp}, Hideyuki Umeda$^{1}$, Keiichi Maeda$^{2,3}$, and Tatsuo Ishii$^{4}$ \\
$^{1}$Department of Astronomy, Graduate School of Science, University of Tokyo, 
Tokyo 113-0033, Japan\\
$^{2}$Department of Astronomy, Graduate School of Science, Kyoto University, Kyoto 606-8502, Japan\\
$^{3}$Kavli Institute for the Physics and Mathematics of the Universe (WPI), University of Tokyo, 
Chiba 277-8583, Japan\\
$^{4}$Department of General Systems Studies, Graduate School of Arts and Sciences, University of Tokyo, Tokyo 153-8902, Japan\\
}
\begin{document}

\date{Accepted 2015 December 22, Received 2015 December 11; in original form 2015 November 5}

\pagerange{\pageref{firstpage}--\pageref{lastpage}} \pubyear{2016}

\maketitle

\label{firstpage}

\begin{abstract}
Massive stars having a CO core of $\sim$40--60 M$_\odot$ experience 
pulsational pair-instability (PPI) after carbon-burning.
This instability induces strong pulsations of the whole star and a part of outer envelope
is ejected.
We investigate the evolution and mass ejection of metal-poor very massive stars which
experience PPI.
We use stellar models with initial masses of 140, 200, and 250 M$_\odot$ and the metallicity $Z=0.004$.
Their masses decrease to 54.09, 58.65, and 61.03 M$_\odot$ before the neon-burning
owing to mass-loss and He mass fraction at the surface becomes about 20 per cent.
During the PPI period of $\sim$1--2000 yr, they experience six, four, and three pulsations,
respectively.
The larger CO-core model has the longer PPI period and ejects the larger amount of mass.
Since almost all surface He has been lost by the pulsations, these stars become
Type Ic supernovae if they explode.
Light curves during the PPI stage and supernovae are investigated and are implicated in luminous 
supernovae.
The luminosity created by the interaction of different PPI ejecta becomes
$M_{{\rm bol}} \sim -16$ to $-20$.
The interaction between the circumstellar shell ejected by PPI and the supernova ejecta can be 
more luminous.
These luminous transients could be an origin of Type I superluminous supernovae and
supernovae with precursor.
\end{abstract}


\section{Introduction}

Massive stars in a certain mass range experience strong pulsations by an instability induced by
electron-positron pair-creation during O- and Si-burning.
This pulsation event is called as pulsational pair-instability (PPI).
The pulsation mechanism is the same as pair-instability supernova (PISN) but the released
energy is too small to explode the whole star.
The mass range of the corresponding CO core is about 40--60 M$_\odot$
\citep{Heger02, Umeda02, Umeda08, Waldman08}, which is smaller than that of the PISN
progenitors, $\sim$60--130 M$_\odot$ \citep[e.g.][]{Heger02, Umeda02, Takahashi15}.
In metal-poor stars, the mass range strongly depends on mass-loss rate.
A part of very massive stars with the metallicity $Z \la 0.004$ experience the PPI
\citep{Yoshida11, Yoshida14}.
Recently, evolution of metal-poor very massive stars has been investigated 
\citep[e.g.][]{Langer07, Yungelson08, Yoshida11, Yusof13, Yoshida14} for 
superluminous supernovae (SLSNe) \citep{Smith07, Gal-Yam09} and 
new findings of very massive stars \citep{Crowther10}.

Strong pulsations by the PPI induce eruptive mass-loss.
The mass ejection from the H-rich envelope and He-layer has been investigated 
in \citet{Woosley07}.
The case of the first pulsation in rotating He stars and CO stars has 
been examined in \citet{Chatzopoulos12}.
The collisions of the ejected matter to the earlier ejecta induce brilliant optical transients.
The brilliant events by the collision during the PPI have been numerically shown in 
\citet{Woosley07}.
The luminosity of the evaluated light curve can be similar to SLSNe.

SLSNe have recently been found as very bright supernovae (SNe) 
\citep[$M_{\rm peak} < -21$;][]{Gal-Yam12}.
SLSNe of Type I and Type II (SLSNe-I and SLSNe-II) were identified as SNe Ic and SNe IIn, 
respectively, and their light curves have large diversity and some of them indicate fast declines 
\citep[e.g.][]{Quimby07, Quimby11}.
SLSNe-R show slow decline and are considered to be powered by the radioactive decays
of $^{56}$Ni and $^{56}$Co \citep{Gal-Yam09, Gal-Yam12}.
However, emission processes and explosion mechanism of these SNe are not still clarified.
Several possibilities of emission processes such as
the interaction between circumstellar medium (CSM) released by eruptive mass-loss 
and the SN ejecta, mass ejection from nascent magnetars \citep[e.g.][]{km2007, kasen2010}, 
PISNe \citep{Gal-Yam09}, 
and core-collapse (CC) SNe after PPI \citep{Moriya10, Yoshida14} have been proposed.
Another class of candidates of PPI events are bright optical transients which were found 
in a few years before SN explosion
for some SNe such as SN 2006jc \citep{Pastorello07} and SN 2009ip \citep{Pastorello13}.
The transients would be powered by outbursts, and a part of the envelope would have been
removed.
This would be related to the PPI of very massive stars.

We investigate the mass ejection by the PPI of massive Wolf--Rayet (WR) stars evolved from
metal-poor very massive stars.
We also investigate the time evolution of resulting optical events during the PPI stage and 
SN explosions of massive WR stars.
The massive WR stars are more compact than red giants having H-rich envelope.
The surface composition of these stars mainly consists of C and O, and a small amount of He
is also included.
Therefore, these transients and SN explosions will be observed as SNe Ib or SNe Ic.
We note that whether these WR stars explode as SNe has not been clarified.
If these stars collapse and do not explode, the luminous optical event caused by SNe
will not be seen.

We organize this paper as follows.
We describe the models for stellar evolution and the mass ejection by PPI in Section 2.
Then, we show details for the stellar evolution and mass ejection during the PPI stage 
of 140, 200, and 250 M$_\odot$ models.
We briefly show the characteristics of their SN explosions with the explosion
energy of $10^{51}$ erg and $10^{52}$ erg in Section 3.
In Section 4, we show light curves of optical transients during the PPI stage and SN explosions.
We discuss characteristics of PPI in our models in Section 5.
We also discuss observations of the optical transients with precursors and SLSNe 
from a viewpoint of the PPI and SN explosions of very massive stars.
We summarize this study in Section 6.

\section{Stellar Evolution and mass ejection during PPI stage}

We pursue the evolution and mass ejection during PPI stage of 140, 200, and 250 star models 
using a stellar evolution code and a hydrodynamic code.
First, we describe details for the calculation method in this stage.
Then, we show the results of the evolution and mass ejection.
These stars pulsate several times by pair instability and eject their outer region by each pulsation.
The total mass and surface composition change by the pulsations.
We also present the stellar mass dependence of detailed evolution.

\begin{table}
\caption{
The total mass $M_{{\rm total}}$, CO core mass $M_{{\rm CO}}$, and the surface mass fractions 
of He ($X_{\rm S}$(He)), C ($X_{\rm S}$(C)), and O ($X_{\rm S}$(O)) of the stellar models.
}
%\begin{center}
\begin{tabular}{lccccc}
\hline
$M_{{\rm ZAMS}}$ & $M_{{\rm total}}$ & $M_{{\rm CO}}$ & $X_{\rm S}$(He) & $X_{\rm S}$(C) & 
$X_{\rm S}$(O) \\ 
(M$_\odot$) & (M$_\odot$)  & (M$_\odot$) & & &  \\
\hline
140 & 54.09 & 48.09 & 0.192 & 0.415 & 0.386 \\
200 & 58.65 & 52.28 & 0.198 & 0.412 & 0.384 \\
250 & 61.03 & 53.95 & 0.193 & 0.407 & 0.393  \\
\hline
\end{tabular}
%\end{center}
\end{table}

\begin{figure}
\includegraphics[width=6cm,angle=270]{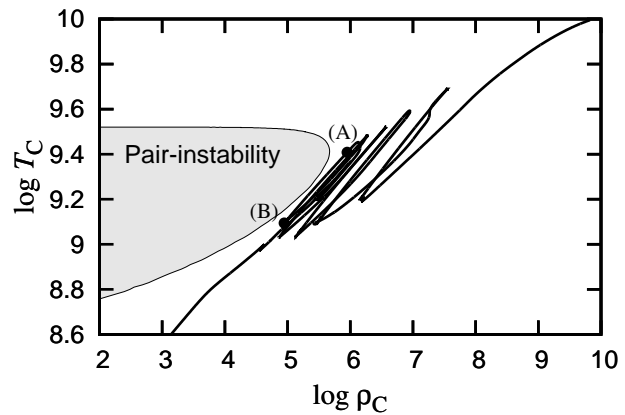}
\caption{The evolution of the 140 M$_\odot$ model on the plane of 
$\log \rho_{\rm C} - \log T_{\rm C}$
after He core burning.
The shaded region shows unstable region against electron--positron pair instability.
Details for points (A) and (B) are explained in text.
}
\end{figure}

\subsection{Calculation method of stellar evolution and mass ejection during PPI stage}

In order to calculate advanced stages of very massive stars, we use the stellar structure
calculated in \citet{Yoshida14}.
We take three models with the initial masses of 140, 200, and 250 M$_\odot$ and the 
metallicity of $Z=0.004$ as the initial structure.
These stars lose H envelope and He layer and evolve to WO stars.
The total mass, CO core mass, and surface composition after the C-burning are listed
in Table 1.
The CO-core mass is defined as the largest mass coordinate satisfying He mass fraction 
smaller than $10^{-3}$.
We calculate the evolution after the C burning to the onset of the CC using
the stellar evolution code adopted in \citet{Yoshida14}.
The acceleration term is included in this code.
Note that exact expression of energy generation in the formulation of stellar evolution 
is discussed in \citet{Takahashi15}.
This expression has not been included in this study.

We calculate the evolution during PPI stage using the stellar evolution code
and a {\sc ppm} hydrodynamic code \citep[e.g.][]{Colella84, Umeda05} as follows.
This is because it is difficult to solve unbound surface structure during the PPI mass ejection 
using the stellar evolution code.

\begin{figure}
\includegraphics[width=6cm,angle=270]{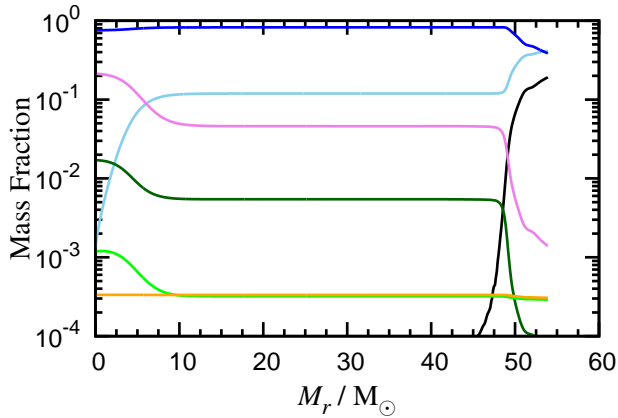}
\caption{Mass fraction distribution of the 140 M$_\odot$ model after the C-burning
at $\log T_{\rm C} = 9.14$.
Black, sky-blue, blue, purple, dark-green, green, and orange lines indicate 
the mass fractions of
$^4$He, $^{12}$C, $^{16}$O, $^{20}$Ne, $^{24}$Mg, intermediate-mass elements 
(Si-Sc), and Fe-peak elements (Ti-), respectively.
}
\end{figure}

\begin{figure}
\includegraphics[width=6cm,angle=270]{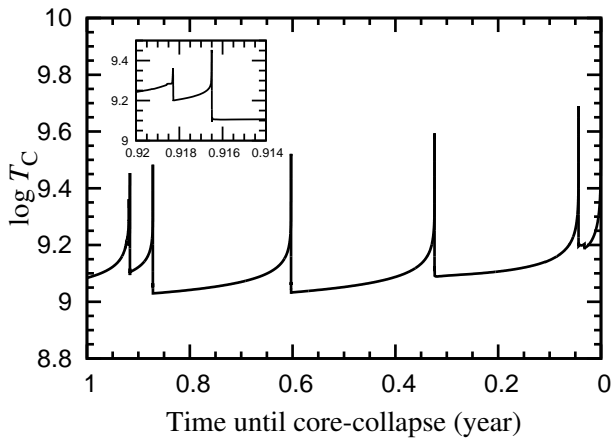}
\caption{Time evolution of the central temperature $\log T_{\rm C}$ of the 140 M$_\odot$ model 
during PPI stage.
The small figure shows the central temperature between 0.92 yr and 0.914 yr.
}
\end{figure}

We describe the calculation method using the second pulsation of the 140 M$_\odot$ star
model as an example.
Fig. 1 shows the evolutionary track of the 140 M$_\odot$ star model on the plane of
$\log \rho_{\rm C}$ (the central density) and $\log T_{\rm C}$ (the central temperature).
Two points (A) and (B) in the second pulsation are assigned.
First, we calculate the stellar evolution passing through points (A) and (B) without mass-loss
by the PPI using the stellar evolution code.

Next, using the {\sc ppm} code, we calculate the mass ejection of the star for about $10^4$ s, 
corresponding to the period until the mass ejection is almost completed.
The initial structure is set to be the structure at the time 
when the star has the maximum total energy during the pulsation (point (A) in the second
pulsation).
During the expansion, outermost region of the star accelerates and exceeds escape velocity.
Thus, the mass exceeding the escape velocity is considered as the ejected mass.

Then, we restart the stellar evolution calculation
from the time when the central temperature is almost minimal (point (B) in the second pulsation).
We artificially reduce stellar mass with a constant mass-loss rate evaluated in the 
previous hydrodynamic calculation. 
The mass-loss continues until the lost mass is equal to the ejected mass obtained from
the hydrodynamic calculation.
After several pulsations, the star forms an Fe core and collapses.
We stop the calculation of the stellar evolution when the central temperature
reaches $10^{10}$K.

\subsection{Stellar evolution during PPI stage}

First, we show the evolution of the 140 M$_\odot$ model after the C-burning.
Fig. 2 shows the mass fraction distribution after the C-burning.
The central temperature is $\log T_{\rm C} = 9.14$.
The main component at the centre is O, Ne, and Mg.
The mass fraction of Si also increases.
The central C burning occurs radiatively.
The O/C/Ne layer is in the range of $\sim$6--48 M$_\odot$ in the mass coordinate.
The surface component is mainly O and C, and 19 per cent $^4$He by mass fraction remains
as shown in Table 1.

The time evolution on the plane of the central temperature and density is shown in Fig. 1.
This star pulsates six times by pair instability.
The evolution line passes close to the right of the pair-instability region.
After the temperature minimal in each pulse,
the slope of the evolution line becomes less steep.
In this stage, the star contracts quasi-statically and neutrino cooling becomes more effective.
As the contraction timescale becomes shorter, the evolution line becomes steeper.

\begin{figure*}
\includegraphics[width=4cm,angle=270]{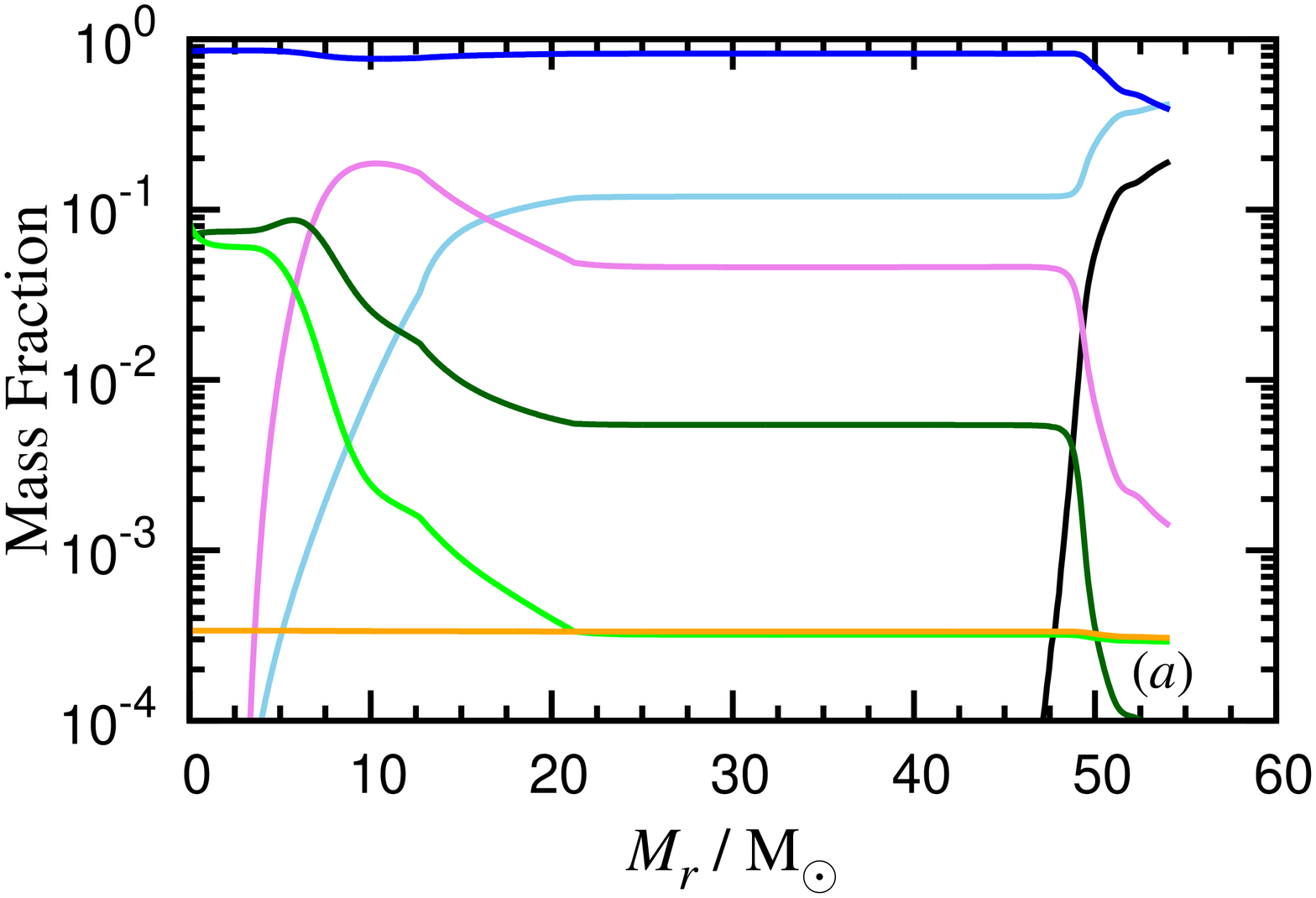}
\includegraphics[width=4cm,angle=270]{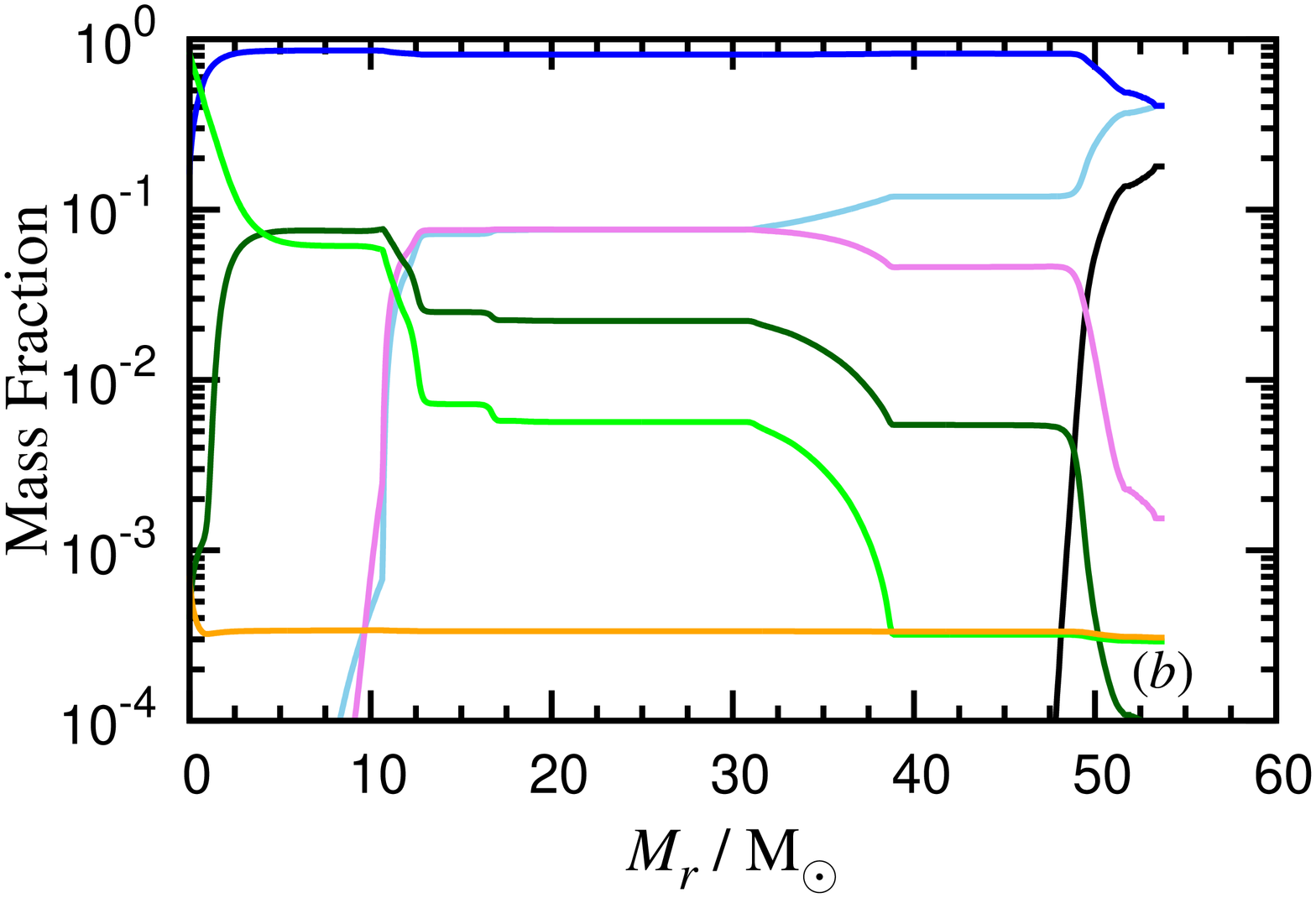}
\includegraphics[width=4cm,angle=270]{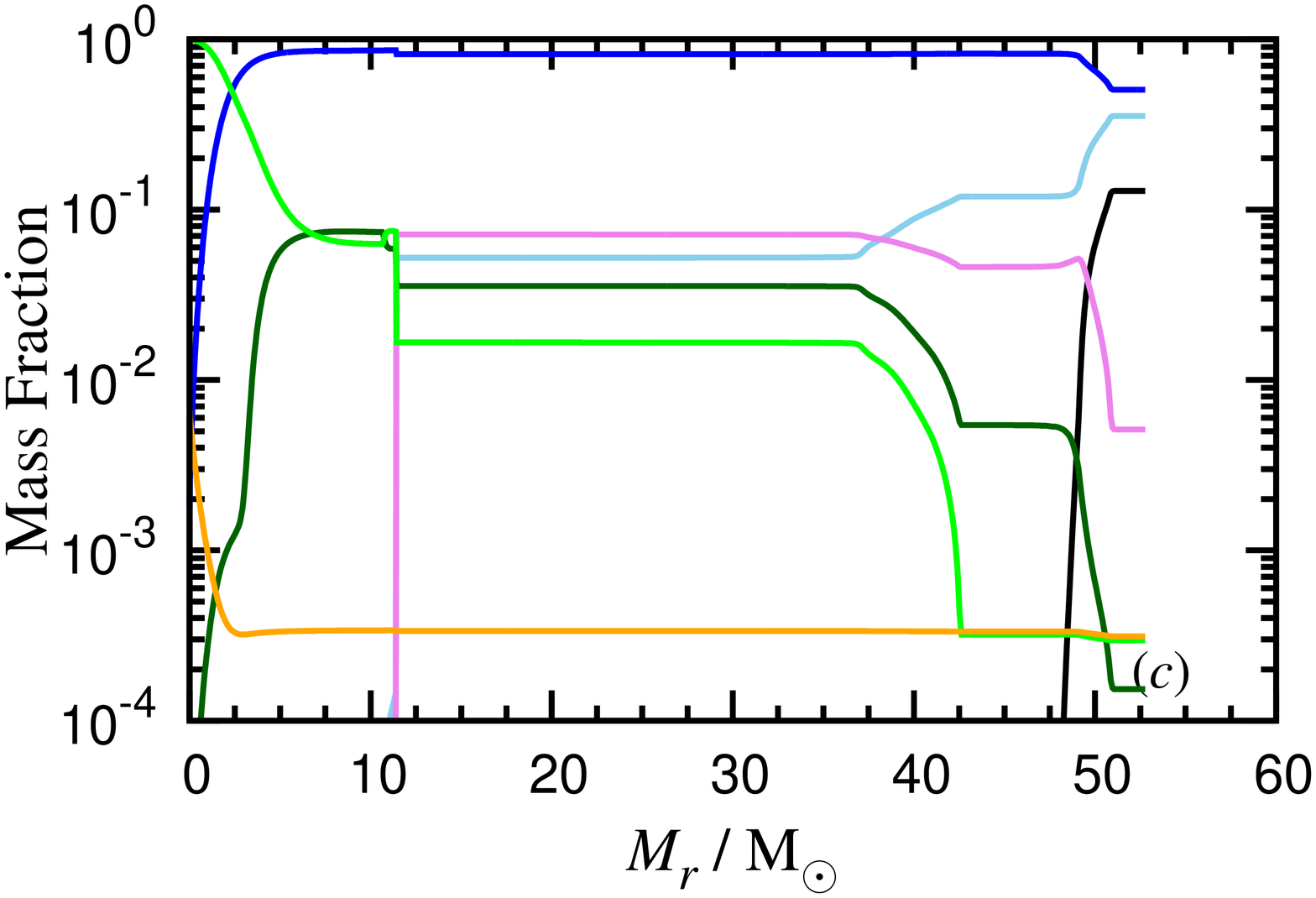}
\includegraphics[width=4cm,angle=270]{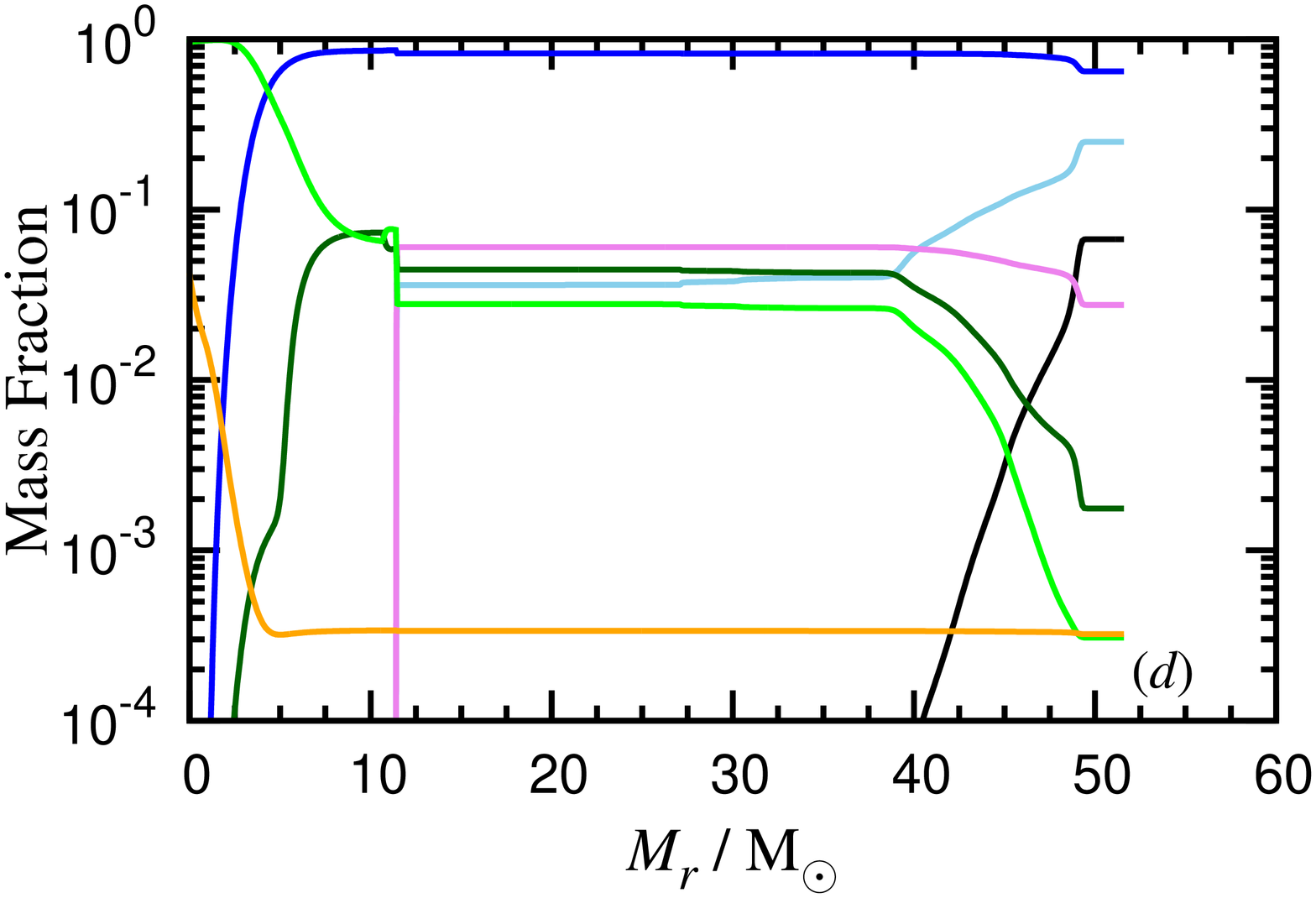}
\includegraphics[width=4cm,angle=270]{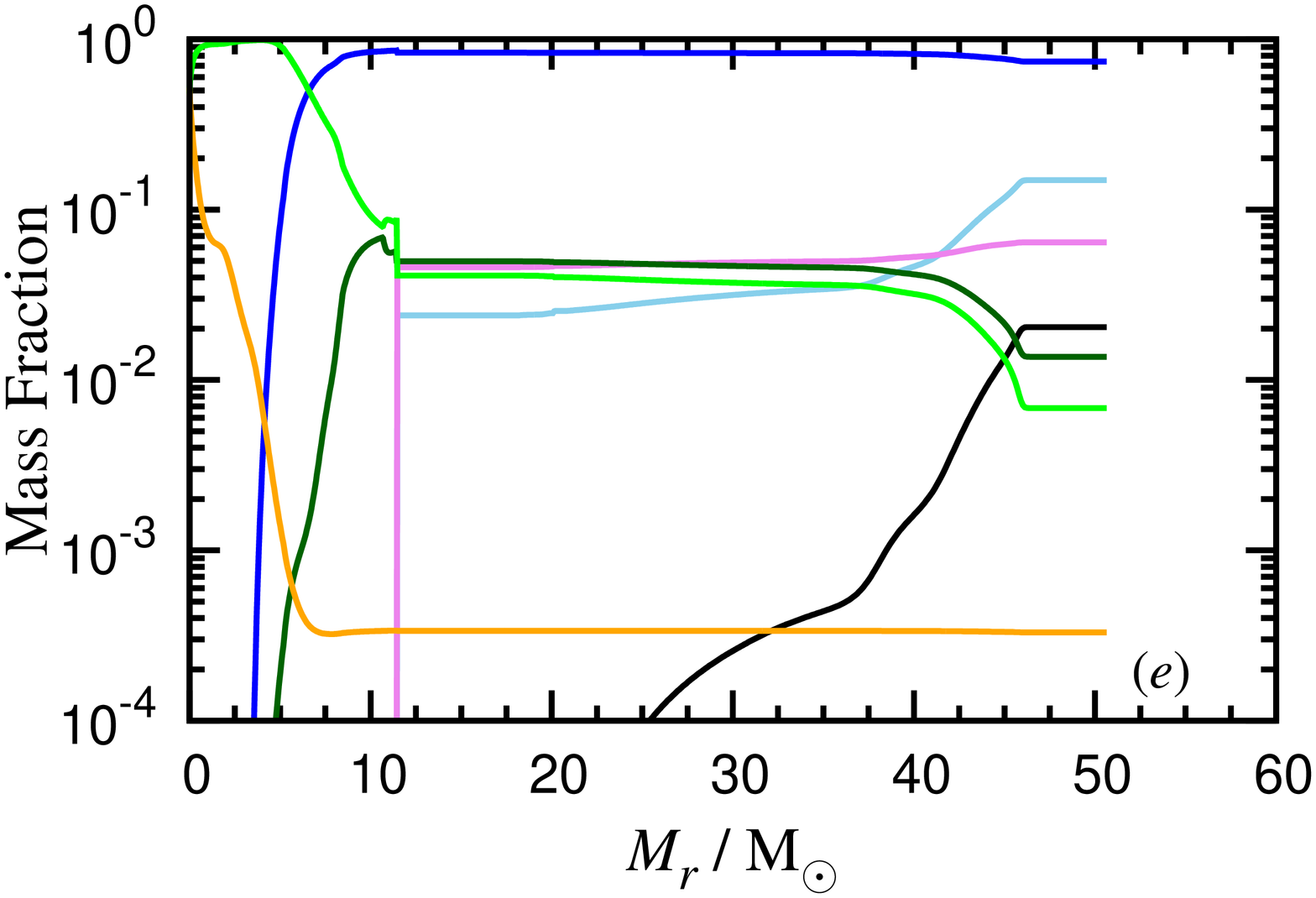}
\includegraphics[width=4cm,angle=270]{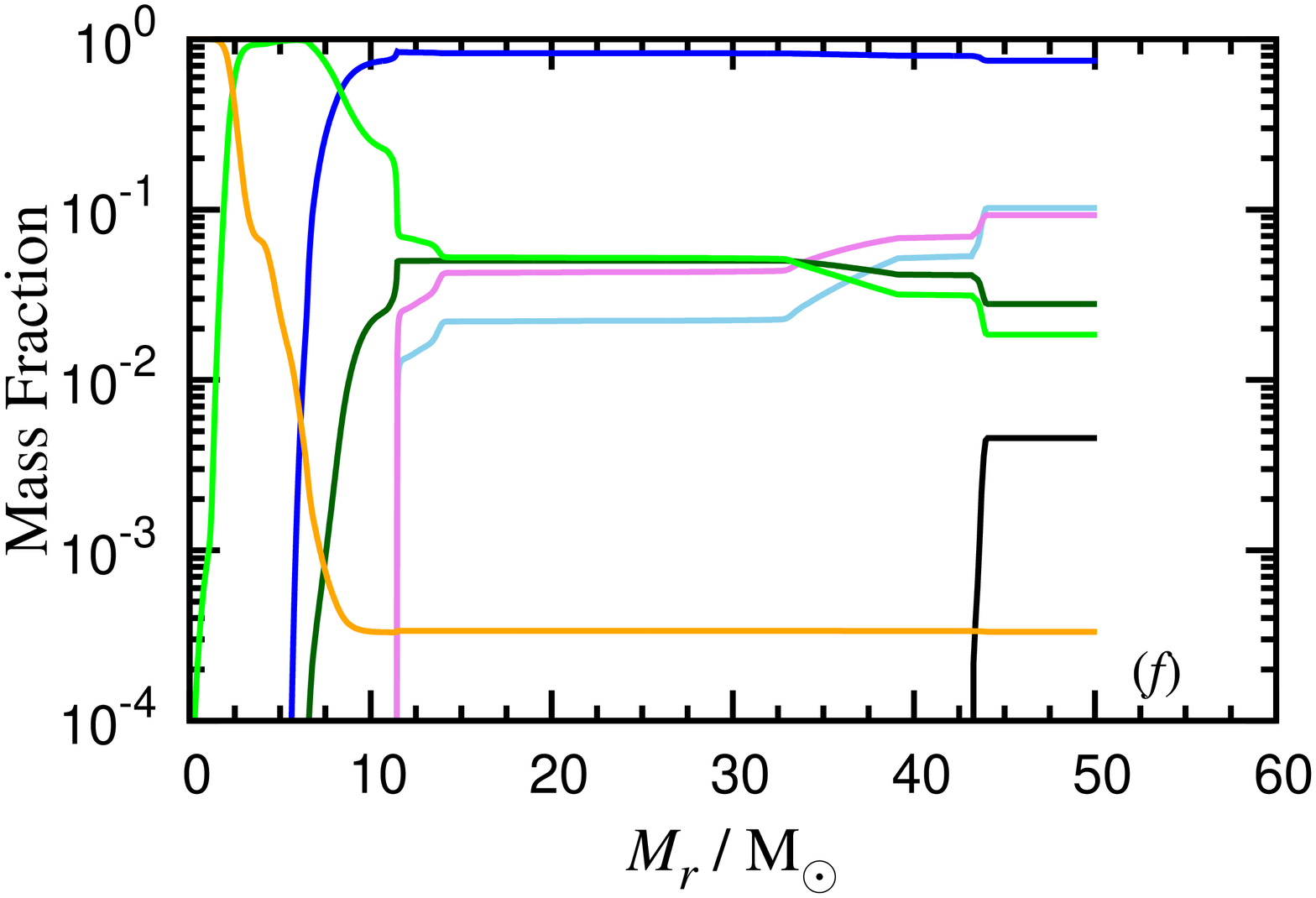}
\caption{Mass fraction distribution of the 140 M$_\odot$ model after the first (a),  
second (b), third (c), fourth (d), fifth (e),  and sixth (f) pulses.
Black, sky-blue, blue, purple, dark-green, green, and orange lines indicate the mass fractions of
$^4$He, $^{12}$C, $^{16}$O, $^{20}$Ne, $^{24}$Mg, Si-Sc, and Fe-peak elements (Ti-),
respectively.
}
\end{figure*}
Fig. 3 shows the time evolution of the central temperature.
The ejected mass and the kinetic energy of the ejecta in each pulsation are shown in Table 2.
The PPI period, i.e., the period from the central carbon exhaustion to the onset of  the CC, is about 1 yr.
After 6.2 d from the central carbon exhaustion, the contraction changes to dynamical.
The dynamical contraction increases the temperature in a short timescale 
and causes the central Ne/O-burning and shell C-burning.
This contraction stops at the central temperature of $\log T_{\rm C} = 9.36$ and the star
expands by rapid Ne/O-burning.
The expansion continues until $\log T_{\rm C} = 9.20$.
Fig. 4 (a) shows the mass fraction distribution after the 1st pulsation.
The mass fractions of $^{16}$O, $^{24}$Mg, and  $^{28}$Si are 0.84, 0.07, and 0.07 at the centre, respectively.
The O/Ne layer forms in $M_{\rm r} \sim 6.7$--16 M$_\odot$.
The chemical composition outside $\sim 20$ M$_\odot$ is not affected by this pulsation.
This expansion is too weak to erupt the CO-rich envelope.
Then, the next contraction proceeds for 15 hr.

\begin{figure}
\includegraphics[width=6cm,angle=270]{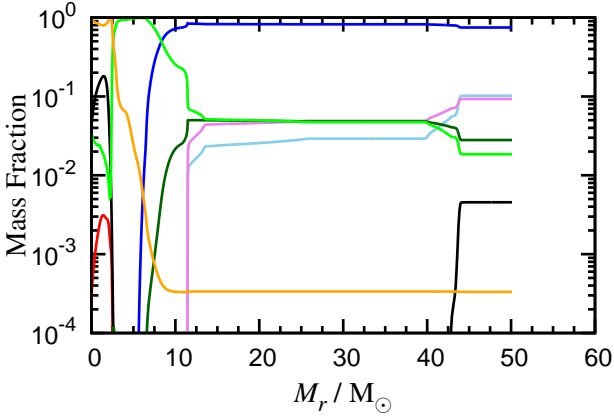}
\caption{Mass fraction distribution of the 140 M$_\odot$ model at the final time step.
Red, black, sky-blue, blue, purple, dark-green, green, and orange lines indicate the mass fractions of
$^1$H, $^4$He, $^{12}$C, $^{16}$O, $^{20}$Ne, $^{24}$Mg, Si-Sc, and Fe-peak elements (Ti-),
respectively.
}
\end{figure}

\begin{table}
\caption{
The mass of the star or remnant $M$, ejecta mass $M_{{\rm ej}}$, 
kinetic energy of ejecta $E_{{\rm kin}}$, 
and time interval to the next pulse $\Delta t_{{\rm pulse}}$ or
SN explosions for each pulse of the 140, 200, and 250 M$_\odot$ models.
}
\begin{tabular}{lcccc}
\hline
Pulse or SN & $M$ & $M_{{\rm ej}}$ & $E_{{\rm kin}}$ & $\Delta t_{{\rm pulse}}$ \\
 & (M$_\odot$) & (M$_\odot$) & (erg) & (yr) \\
\hline
\multicolumn{5}{c}{140 M$_\odot$ model ($M = 54.09$ M$_\odot$ before PPI)} \\
\hline
First pulse & 54.09 & 0.00 & 0.00 & $1.77 \times 10^{-3}$ \\
Second pulse & 53.84 & 0.25 & $1.26 \times 10^{49}$ & $4.69 \times 10^{-2}$ \\
Third pulse & 52.77 & 1.07 & $3.74 \times 10^{49}$ & $2.69 \times 10^{-1}$ \\
Fourth pulse & 51.59 & 1.18 & $3.83 \times 10^{49}$ & $2.79 \times 10^{-1}$ \\
Fifth pulse & 50.65 & 0.94 & $2.48 \times 10^{49}$ & $2.79 \times 10^{-1}$ \\
Sixth pulse & 50.10 & 0.55 & $2.99 \times 10^{49}$ & $4.32 \times 10^{-2}$ \\ 
SN ($10^{51}$ erg) & 39.49 & 10.61 & $1.02 \times 10^{51}$ & \\
SN ($10^{52}$ erg) & 1.95 & 48.12 & $9.71 \times 10^{51}$ & \\
\hline
\multicolumn{5}{c}{200 M$_\odot$ model ($M = 58.65$ M$_\odot$ before PPI)} \\
\hline
First pulse & 57.13 & 1.52 & $6.39 \times 10^{49}$ & $9.05 \times 10^{-1}$ \\
Second pulse & 55.08 & 2.05 & $5.62 \times 10^{49}$ & $2.03 \times 10^{1}$ \\
Third pulse & 54.38 & 0.70 & $1.93 \times 10^{49}$ & $5.34 \times 10^{-2}$ \\
Fourth pulse & 53.35 & 1.03 & $6.53 \times 10^{49}$ & $8.52 \times 10^{-2}$ \\
SN ($10^{51}$ erg) & 41.82 & 11.53 & $9.97 \times 10^{50}$ & \\
SN ($10^{52}$ erg) & 1.96 & 51.39 & $1.03 \times 10^{52}$ & \\
\hline
\multicolumn{5}{c}{250 M$_\odot$ model ($M = 61.03$ M$_\odot$ before PPI)} \\
\hline
First pulse & 60.48 & 0.55 & $2.90 \times 10^{49}$ & $1.10 \times 10^{-1}$ \\
Second pulse & 56.26 & 4.22 & $3.89 \times 10^{50}$ & $1.43 \times 10^{3}$ \\
Third pulse & 53.16 & 3.10 & $2.87 \times 10^{50}$ & 3.74 \\
SN ($10^{51}$ erg) & 38.64 & 14.52 & $1.00 \times 10^{51}$ & \\
SN ($10^{52}$ erg) & 1.98 & 51.779 & $1.08 \times 10^{52}$ & \\
\hline
\end{tabular}
\end{table}

The second pulsation raises the central temperature up to $\log T_{\rm C} = 9.45$.
Then, the central temperature decreases to $\log T_{\rm C} = 9.09$ by the expansion.
The central $^{16}$O is burned to intermediate elements such as Si and S.
The mass fraction of $^{16}$O at the centre decreases to 0.16 (see Fig. 4 (b)).
The O/Mg/Si layer is formed in the region inside about 10 M$_\odot$.
A part of the O/Ne layer and the outer O/Si layer mix, thus, the composition
in $M_{\rm r} \la 30$ M$_\odot$ also changes. 
The mass fractions of Ne and C in this region are roughly equal.
In this pulsation, the outer region of 0.25 M$_\odot$ is ejected through the rapid
expansion.
The quasi-static period in the next contraction is about 16 d.

In the third pulsation, the central temperature rises up to $\log T_{\rm C} = 9.48$.
After the expansion, the central temperature decreases to $\log T_{\rm C} = 9.03$.
The O mass fraction at the centre decreased to $2 \times 10^{-3}$ and the main component
becomes intermediate nuclei such as Si and S (see Fig. 4 (c)).
The outer boundary of the surrounding O/Mg/Si layer becomes about 11 M$_\odot$.
In the region inside 38 M$_\odot$, the Ne mass fraction becomes larger
than C owing to shell C-burning and the mixing.
This expansion reduces the stellar mass by 1.07 M$_\odot$.
The surface He mass fraction decreases to 0.13.
The quasi-static contraction after the expansion continues for 98 d.

In the fourth pulsation, the central temperature reaches $\log T_{{\rm C}} = 9.52$.
Then, the expansion decreases to the temperature to $\log T_{\rm C} = 9.03$.
The central O-burning proceeds so that the Si core grows up to 4.3 M$_\odot$
(see Fig. 4 (d)).
This pulsation also expels the surface of 1.18 M$_\odot$.
The surface He mass fraction decreases to 0.07.
The next quasi-static contraction continues for 102 d.

\begin{table*}
\caption{
The final mass $M_{\rm f}$, Fe-core mass $M_{\rm Fe}$, CO-core mass $M_{{\rm CO}}$, 
the surface mass fractions of $^{4}$He $X_S$($^4$He), $^{12}$C $X_S$($^{12}$C), 
$^{16}$O $X_S$($^{16}$O), $^{20}$Ne $X_S$($^{20}$Ne), $^{24}$Mg $X_S$($^{24}$Mg), 
and $^{28}$Si $X_S$($^{28}$Si) at the final stage.
}
\begin{tabular}{lccccccccc}
\hline
Model & $M_{\rm f}$/M$_\odot$ & $M_{{\rm Fe}}$/M$_\odot$ & $M_{{\rm CO}}$/M$_\odot$ & 
$X_S$($^4$He) & $X_S$($^{12}$C) & $X_S$($^{16}$O) & $X_S$($^{20}$Ne) & 
$X_S$($^{24}$Mg) & $X_S$($^{28}$Si)  \\
\hline
140 & 50.10 & 2.57 & 43.51 & $4.5 \times 10^{-3}$ & 0.102 & 0.748 & 0.093 & 0.028 & 0.018 \\
200 & 53.35 & 2.38 & 53.35 & $3.3 \times 10^{-5}$ & 0.084 & 0.765 & 0.093 & 0.036 & 0.013 \\
250 & 53.16 & 2.10 & 53.16 & $1.9 \times 10^{-5}$ & 0.046 & 0.812 & 0.064 & 0.043 & 0.028 \\
\hline
\end{tabular}
\end{table*}

\begin{figure}
\includegraphics[width=6cm,angle=270]{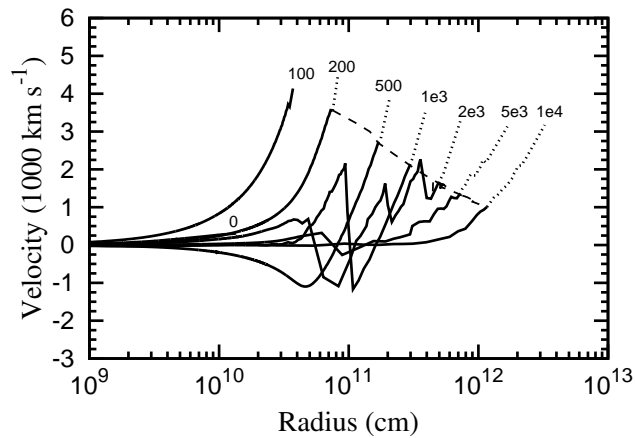}
\caption{The time variation of the velocity distribution as a function of the radius during the
fourth pulsation of the 140 M$_\odot$ model.
In each line, solid part indicates the gravitationally bounded region and dotted part indicates
the unbound region, i.e., the velocity exceeds the escape velocity.
The dashed line indicates the outer boundary of the bounded region.
The number attached to each line denotes the time after the star has the maximum total energy 
in the pulsation. 
}
\end{figure}

\begin{figure}
\includegraphics[width=6cm,angle=270]{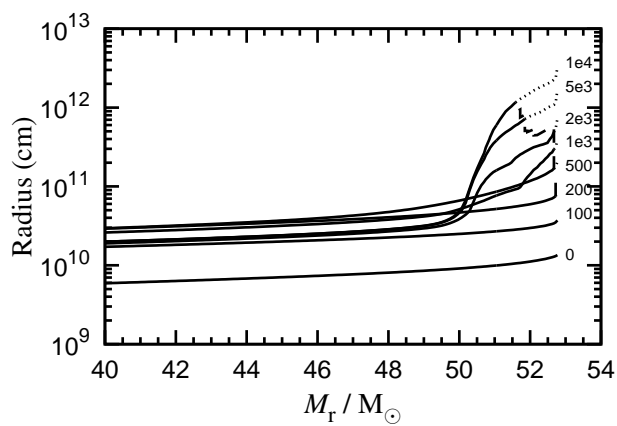}
\caption{The time variation of the radius distribution as a function of the mass coordinate during the
fourth pulsation of the 140 M$_\odot$ model.
In each line, solid part indicates the gravitationally bounded region and dotted part indicates
the unbound region.%, i.e., the velocity exceeds the escape velocity.
The dashed line shows the outer boundary of the bounded region.
The number attached to each line denotes the time after the star has the maximum total energy 
in the pulsation. 
}
\end{figure}

\begin{figure}
\includegraphics[width=6cm,angle=270]{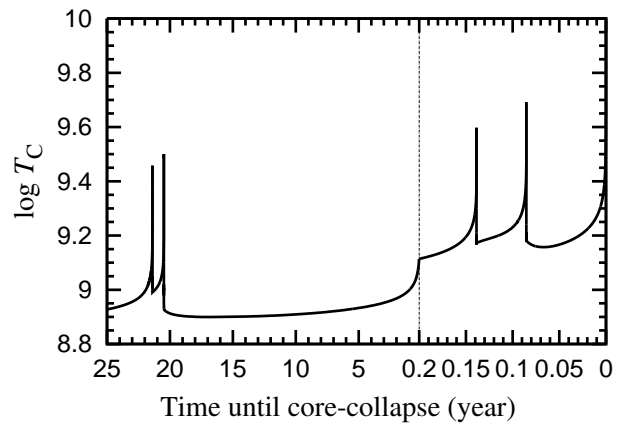}
\caption{Time evolution of the central temperature $\log T_{\rm C}$ of the 200 M$_\odot$ model 
during the PPI stage.
The scale of the horizontal axis changes at 0.2 yr, which is indicated by the vertical dashed line.
}
\end{figure}

The fifth pulsation raises the central temperature up to $\log T_{{\rm C}} = 9.59$.
Then, the expansion decreases the central temperature to $\log T_{\rm C} = 9.09$.
The Si core grows up to 6.6 M$_\odot$ through this pulsation (see Fig. 4 (e)).
At the same time, about a half of the intermediate nuclei at the centre are burned
into Fe-peak elements.
The O- and Si-burnings at the central region expand the star.
The envelope of 0.94 M$_\odot$ is lost by the expansion and He mass fraction
at the surface decreases to 0.02.
The next quasi-static contraction continues for 102 d.

In the sixth pulsation, the central temperature becomes $\log T_{{\rm C}} = 9.68$.
This pulsation expands the star until the central temperature becomes $\log T_{{\rm C}} = 9.20$.
This high temperature induces the core Si-burning and a 2.3 M$_\odot$ Fe core is formed
(see Fig. 4 (f)).
The Si and O/Si layers are also formed outside the core.
This expansion expels the outer region of 0.55 M$_\odot$.
The surface He mass fraction decreases to $4.5 \times 10^{-3}$.
Thus, the surface He mass fraction becomes much smaller than that before the PPI stage.

The next contraction does not change to expansion, and the central core gravitationally collapses.
Quasistatic contraction continues for 16 d and gradually changes to the dynamical collapse.
Fig. 5 is the mass fraction distribution at the final step of the 140 M$_\odot$ model.
The Fe-core mass is 2.6 M$_\odot$ and the Si-layer extends to 8.3 M$_\odot$.
This star loses the outer region of 3.99 M$_\odot$ during the pulsations.
He mass fraction at surface decreases to $4.5 \times 10^{-3}$ and a small amount of He
injects into inner region by the mixing.
The final CO-core mass is 43.51 M$_\odot$.
The final mass, Fe-core mass, CO-core mass, the surface mass fractions of $^4$He, $^{12}$C, 
$^{16}$O, $^{20}$Ne, $^{24}$Mg, and $^{28}$Si are listed in Table 3.
Note that the Fe core mass is defined as the largest mass coordinate satisfying the mass
fraction of Fe-peak elements larger than 0.5.

\begin{figure}
\includegraphics[width=6cm,angle=270]{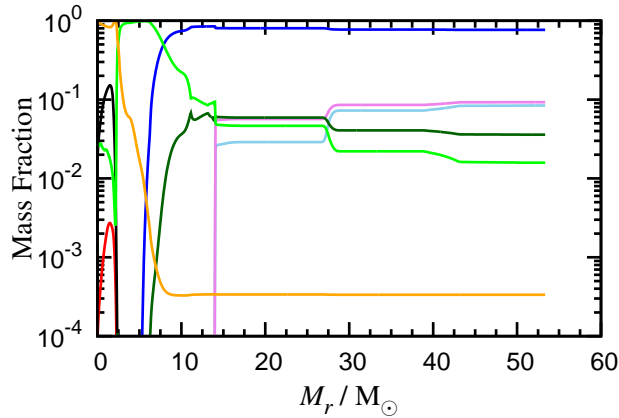}
\caption{Same as Fig. 5, but for the 200 M$_\odot$ model.
}
\end{figure}

\begin{figure}
\includegraphics[width=6cm,angle=270]{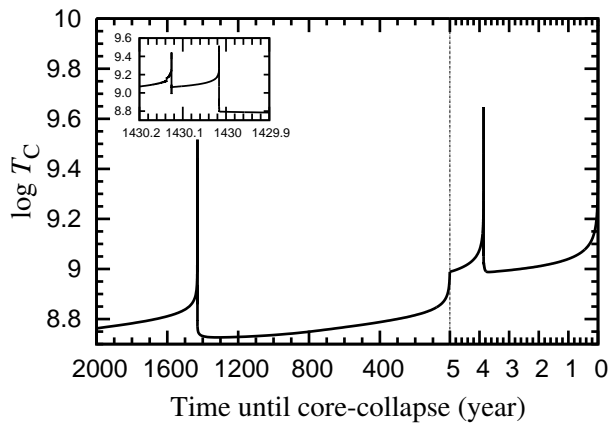}
\caption{Time evolution of the central temperature $\log T_{\rm C}$ of the 250 M$_\odot$ model 
during the PPI stage.
The scale of the horizontal axis changes at 5 yr, which is indicated by the vertical dashed line.
The small figure shows the central temperature between 1430.2 yr and 1429.9 yr.
}
\end{figure}

We explain the mass ejection by the PPI of the 140 M$_\odot$ model.
We present the motion of outermost region during the fourth pulsation of the 140 M$_\odot$
model.
Fig. 6 shows the time evolution of the velocity distribution in the range of $r \ge 10^9$ cm.
First, the velocity in this region is less than 1000 km s$^{-1}$ and the materials
accelerate outwards.
The velocity increases for hundreds  second, and the velocity of the surface material exceeds
the escape velocity at $\sim 200$ s.
Then, the stellar interior starts contracting and the contraction motion gradually propagates
outwards.
However, the surface materials continue to expand.
As a result, the unbound region gradually moves outward.
After several thousand seconds, the unbound surface expands with a constant velocity with 
$\sim 3200$ km s$^{-1}$.
On the other hand, most bounded material contracts slowly.
Finally, the kinetic energy of the ejecta in this pulsation is $3.83 \times 10^{49}$ erg.

Fig. 7 shows the time evolution of the radius as a function of mass coordinate in outer
region ($M_{\rm r} \ge 40$ M$_\odot$) during the fourth pulse.
First, the outer region expands for $\sim 500$ s.
Then, the contraction motion propagates to the outer region and the expanding region
gradually turns into the contraction.
The contraction proceeds very slowly so the radial distribution of the contraction region
changes quasi-statically.
The region with $M_{\rm r} \la 50$ M$_\odot$ does not change the radial distribution after
about 2000 s.
On the other hand, the region with $M_{\rm r} \ga 51$ M$_\odot$ still expands and the outer boundary
of the bound region shifts inward until $\sim 10^4$ s.
Finally, the outer boundary becomes 51.59 M$_\odot$.

\subsection{Stellar mass dependence}

Properties of pulsations induced by PPI depend on the CO-core mass before the 
PPI stage.
Here, we show the properties of the pulsations during PPI stages in the 200 and 250 M$_\odot$
models and compare with the 140 M$_\odot$ model.
Table 2 shows the stellar mass, ejecta mass, kinetic energy of the ejecta, and time interval
from the previous pulse in each pulse.

\subsubsection{200 M$_\odot$ model}

Fig. 8 shows the time evolution of the central temperature of the 200 M$_\odot$ model during
the PPI stage.
This model pulsates four times for about 22 yr before CC.
The contraction period after the second pulsation is 20.3 yr.
This period is much longer than that of other pulsations.
The second pulsation also causes the lowest minimal central-temperature among the pulsations.

The mass ejection of the 200 M$_\odot$ model is also presented in Table 2.
The ejected mass during each pulse is 0.70--2.05 M$_\odot$.
The second pulsation has ejected the largest mass.
The second pulsation also induced the longest quasi-static contraction after the
expansion.
So, this pulsation seems to be the strongest pulsation.
The kinetic energy of the ejecta is (1--4) $\times 10^{49}$ erg in each pulse.
The mass of 5.30 M$_\odot$ is lost by the pulsations and the total mass finally becomes
53.35 M$_\odot$.

\begin{figure}
\includegraphics[width=6cm,angle=270]{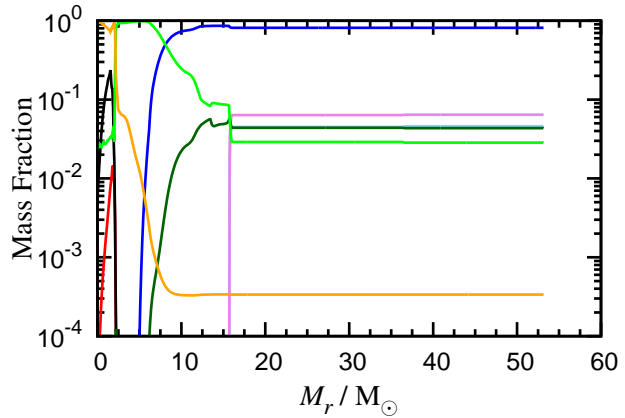}
\caption{Same as Fig. 5, but for 250 M$_\odot$ model.
The line of the $^{12}$C mass fraction in $M_{\rm r} \ge 16$M$_\odot$ almost overlaps 
with that of $^{24}$Mg.
}
\end{figure}

Fig. 9 shows the mass fraction distribution just before the CC.
The Fe core with 2.38 M$_\odot$ forms after the PPI stage.
The surrounding Si-rich layer extends to 7.84 M$_\odot$.
The layer outside the Si-rich layer is O-rich.
The O-rich layer extends from about 4 M$_\odot$.
The second abundant element depends on the mass coordinate.
The mixing during the PPI stage raises the abundances of Si, Mg, and Ne.
The surface He has been lost during the PPI stage and the surface compassion mainly
consists of O, C, and Ne.
He mass fraction at the surface is $3.3 \times 10^{-5}$.

We obtained four pulsations during the PPI stage in the 200 M$_\odot$ model.
The number of pulsations is smaller than the 140 M$_\odot$ model.
On the other hand, the period of the PPI stage is much longer.
The total ejected mass is 5.30 M$_\odot$, which is also larger than that of the 140 M$_\odot$
model.
Thus, the PPI of the 200 M$_\odot$ model is stronger than that of the 140 M$_\odot$ model.
The larger CO core induces stronger contraction by PPI and stronger few pulses can produce
Fe core effectively.

\subsubsection{250 M$_\odot$ model}

Fig. 10 shows the time evolution of the central temperature of the 250 M$_\odot$ model.
This model pulsates three times during the PPI stage.
In the first pulse, the central temperature rises to $\log T_{\rm C} = 9.44$ and more than
half of the central O is burned to Si and other intermediate nuclei.
This pulse ejects the 0.55 M$_\odot$ envelope.
After 0.11 yr from the first pulse, the second pulsation occurs.
The central temperature becomes $\log T_{\rm C} = 9.52$ and the O in the central
region of 0.39 M$_\odot$ is burned.
Then, whole of the star expands and the temperature decreases to $\log T_{\rm C} = 8.73$.
This expansion ejects 4.22 M$_\odot$ outermost region with the kinetic energy of 
$3.9 \times 10^{50}$ erg.
This strong expansion also extends the contraction time of the star.
It takes 1430 yr until the next pulsation.
The third pulse raises the central temperature to $\log T_{\rm C} = 9.65$.
The Fe core with 0.84 M$_\odot$ is formed and 3.10 M$_\odot$ outer region is ejected.

Fig. 11 shows the mass fraction distribution of the 250 M$_\odot$ model at the pre-SN stage.
The Fe-core mass becomes 2.10 M$_\odot$ in this model.
It is smaller than that of the 200 M$_\odot$ model.
The strong expansion would prevent the growth of the Fe core.
There is no He envelope.
He mass fraction at surface is $2 \times 10^{-5}$.
The O/Ne layer extends to the stellar surface.
Almost all He in the outer region is lost during the second pulsation.
The secondary main components are Ne, C, Mg, and Si.
The mixing during the second and third pulsations makes the composition in the O/Ne
layer homogeneous.

The 250 M$_\odot$ model causes a smaller number of pulses, much longer PPI period, 
and larger mass ejection than the 200 M$_\odot$ model.
These trends are opposite to the 140 M$_\odot$ model.
We see from the above results that larger CO core model induces smaller number of pulses,
longer PPI period, and larger mass ejection.
Lower minimal temperature also correlates with a long pulse period.
The large CO core induces stronger contraction.
Stronger contraction causes stronger nuclear burning.
The stronger nuclear burning brings about stronger expansion and larger mass ejection.
It also produces more nuclear-burning product.
The stronger expansion makes the bound of the star looser, so that the period until the
next contraction becomes longer.
Since more nuclear-burning product is synthesized in each pulsation, the number of pulses
to form Fe core enough to collapse is smaller.

\section{SN explosions}

We calculate SN explosions after the CC of the stars using the {\sc ppm} hydrodynamic code.
We inject the explosion energy at  $M_{\rm r} =2.0$ M$_\odot$.
We consider two cases of the explosion energy, $E_{51}$ = 1 and 10, where $E_{51}$ is
the explosion energy in units of $10^{51}$ erg.

The final masses of the central remnant and the ejecta are listed in Table 2.
In the explosion models of $E_{51} = 1$, most of the exploded materials 
do not reach the escape velocity and fall-back to the central remnant.
The ejecta mass is about 10 M$_\odot$, which is much smaller than
the progenitor mass, in the three models.
Although $^{56}$Ni is synthesized in the central region, it falls back to the central remnant
and is not ejected.

In the explosion models with $E_{51} = 10$, all the materials outside the energy-injected region 
are ejected.
We do not include $\alpha$-network in the hydrodynamic calculation.
So, we roughly estimate the ejected amount of $^{56}$Ni, assuming that materials that experience 
temperature higher than $5 \times 10^{9}$ K become entirely $^{56}$Ni.
The obtained $^{56}$Ni mass is 2, 1.5, and 1.2 M$_\odot$ in the 140, 200, 250
M$_\odot$ models, respectively.
In these simulations, we do not adopt the ^^ collapsed-core progenitor model' in 
\citet{Yoshida14} for simplicity.
The $^{56}$Ni mass of  the collapsed-core model would be larger than 
that of the present model by a factor of 2 at most.

\section{Light curves}

\begin{figure*}
\includegraphics[width=5.5cm,angle=0]{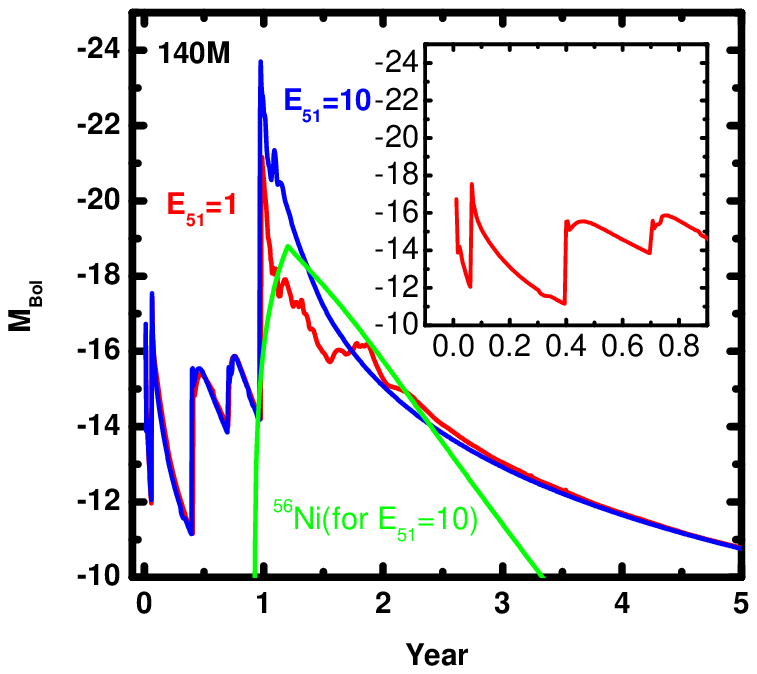}
\includegraphics[width=5.5cm,angle=0]{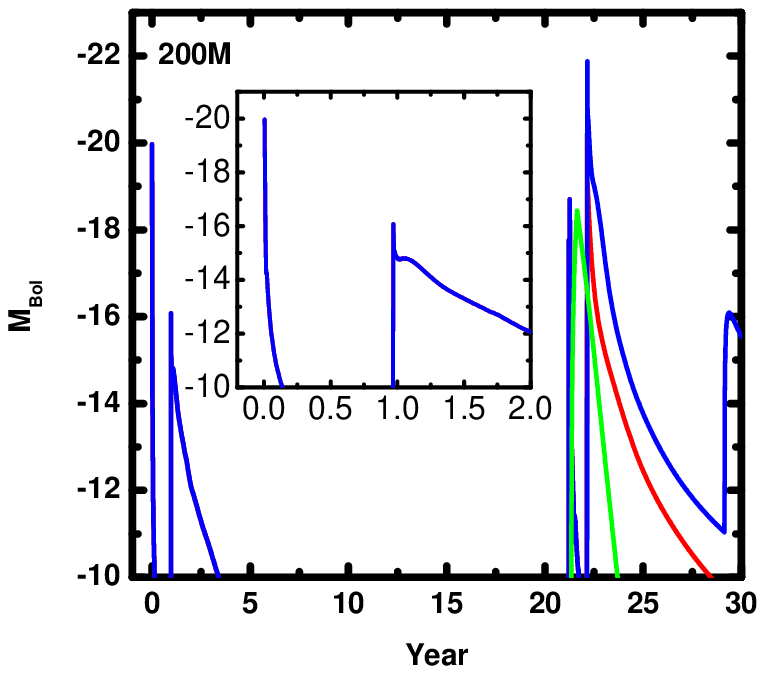}
\includegraphics[width=5.5cm,angle=0]{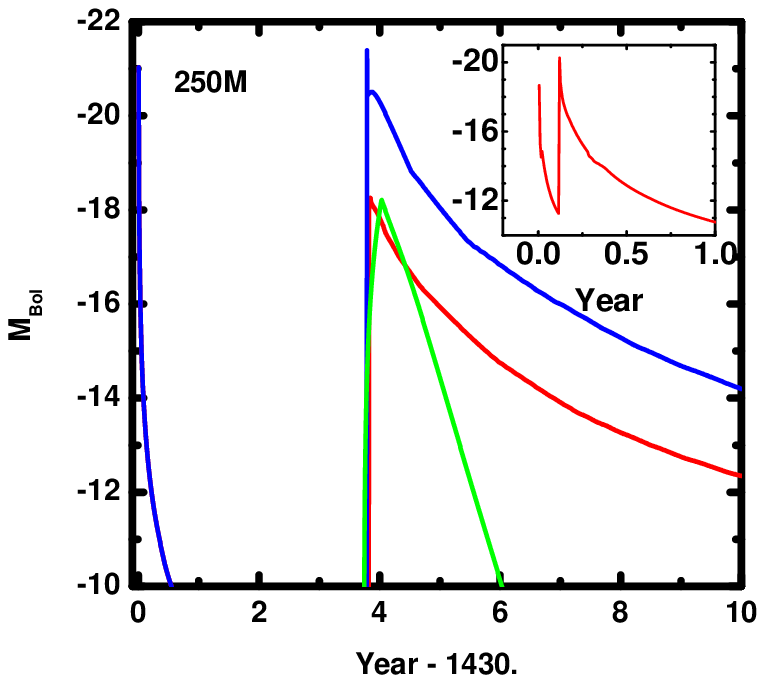}
\caption{Bolometric light curves from the 140 (left), 200 (centre), and 250 M$_{\odot}$ 
(right) models. 
The explosion energy of the final SNe is set to be $E_{51} = 1$ (red) and $10$ (blue). 
In addition, the contribution from the decay of $^{56}$Co, as originally produced as $^{56}$Ni 
at the SN explosion, is indicated (green) for $E_{51} = 10$. 
}
\end{figure*}

\begin{figure*}
\includegraphics[width=5.5cm,angle=0]{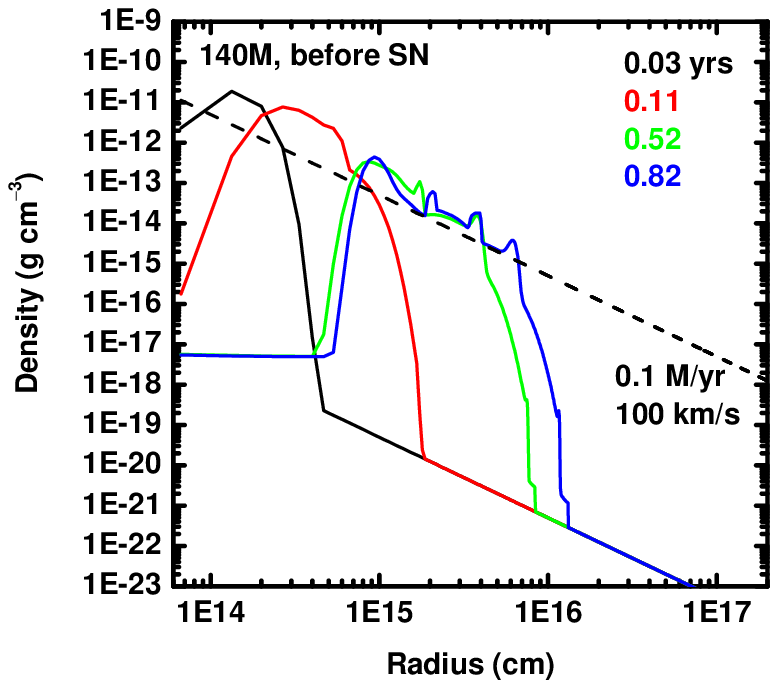}
\includegraphics[width=5.5cm,angle=0]{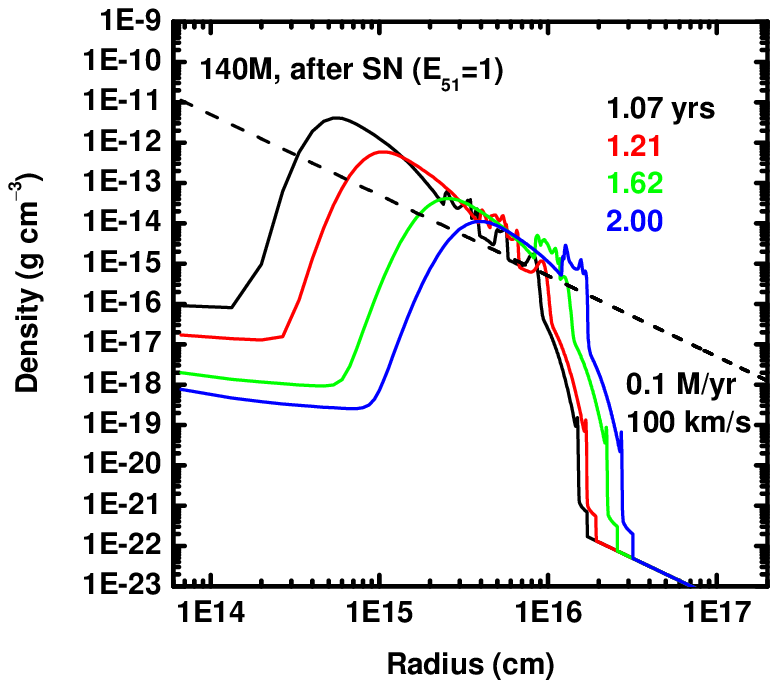}
\includegraphics[width=5.5cm,angle=0]{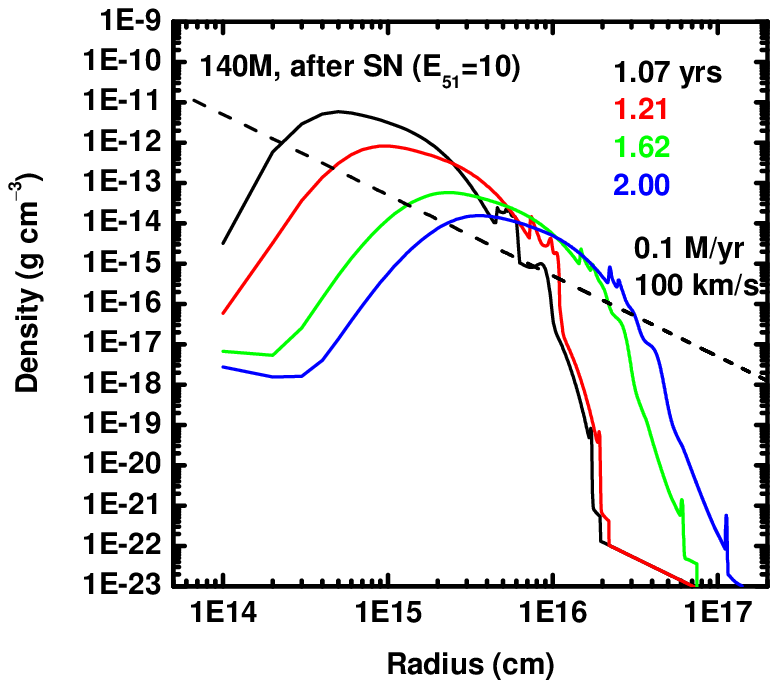}
\caption{Evolution of the density profile for the 140 M$_{\odot}$ model, 
before the SN (left), after the SN with $E_{51}=1$ (centre), and after the SN with
$E_{51}=10$ (right). 
For comparison, the CSM density distribution expected from a steady-state mass-loss of 
0.1 M$_{\odot}$ yr$^{-1}$ with $v_{\rm w} = 100$ km s$^{-1}$, 
a typical value assumed for luminous SNe IIn and SLSNe, is shown by a dashed line.  
}
\end{figure*}

\begin{figure*}
\includegraphics[width=6cm,angle=0]{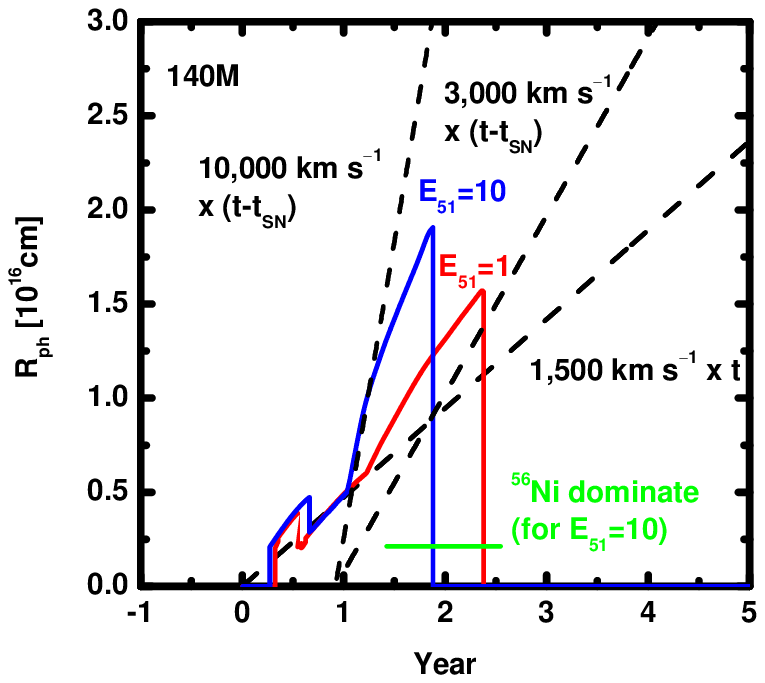}
\includegraphics[width=6cm,angle=0]{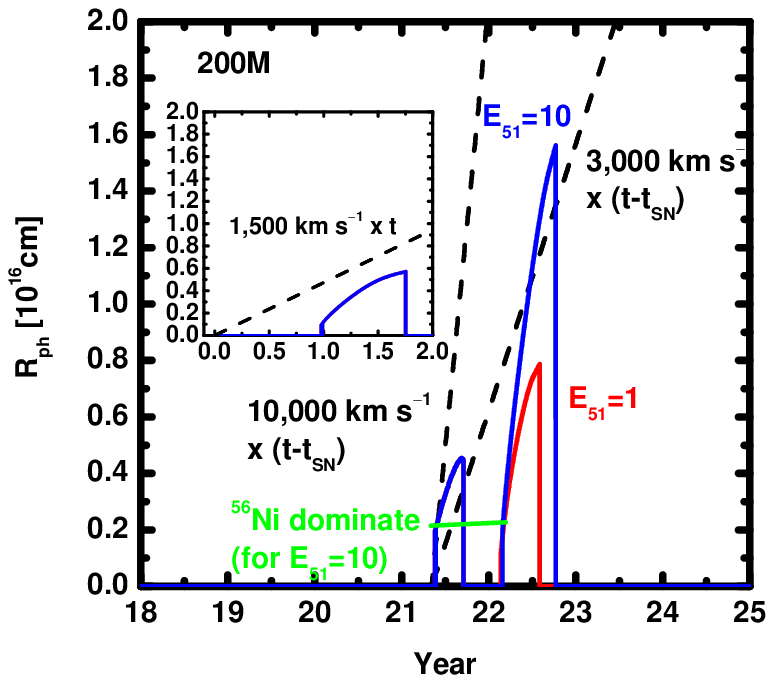}
\caption{Evolution of the photospheric radius as determined by the optical depth 
to electron scatterings in the shocked high-temperature regions, as shown 
for the 140 M$_{\odot}$ (left) and 200 M$_{\odot}$ (right) models. The models with $E_{51} = 1$ are shown by a red line while the models with $E_{51} = 10$ are shown by a blue line. The period during which the $^{56}$Co-decay-powered luminosity exceeds the interaction luminosity is indicated by a green line at the bottom. 
}
\end{figure*}

\subsection{Calculation method}

Adopting the pulsational-mass-loss histories and the final SN explosion models as calculated 
in \S 2.1, we compute bolometric light curves resulting from hydrodynamic interactions between different mass-loss shells and the final SN ejecta. 
The hydrodynamic evolution following the multiple ejections of the shells and SN ejecta 
is followed by a one-dimensional, adiabatic Eulerian hydrodynamic code \citep{km2002,Maeda13}. 
The calculation starts with the injection of the first mass-loss shell, then when the time is elapsed to reach to the ejection of the second mass-loss shell (Table 1), the second shell is injected to the computational domain. This process is repeated until the final SN ejecta are injected to the computational domain.

In the hydrodynamic calculations, we follow evolution of tracer particles which are freshly injected 
with each shell. 
The kinetic energies of all the particles are tracked, which allows us to use equation (7) 
in \citet{Moriya14} to evaluate the kinetic energy which may be dissipated and converted into radiation 
due to the interaction.
We set parameter $\epsilon=0.3$ in equation (8) in \citet{Moriya14} which gives a fraction of the dissipated
energy converted to radiation.

At each time step and position, we estimate the optical depth assuming that the interaction region is fully ionized and that the opacity is dominated by electron scatterings, i.e., 0.2 cm$^{2}$ g$^{-1}$ in absence of hydrogen. This information of the optical depth is used to compute the diffusion time-scale which is then convolved with the energy generation rate in producing the light curve, as well as to compute the position of the photosphere. 
In general, we find that the diffusion time scale is shorter than the dynamical evolution time-scale, and the resulting delay is largely negligible. 
This justifies our procedure to calculate the light curve, where it is assumed that 
the diffusion time-scale is negligible as compared to the dynamical time-scale \citep[see e.g.,][]{Moriya14}.

The above treatment is only for the luminosity created by the hydrodynamic interaction. 
Another source of radiation is $^{56}$Ni synthesized and ejected at the SN explosion. 
For the models with $E_{51} = 1$, virtually no $^{56}$Ni is ejected by the SN, 
and therefore we neglect this contribution. 
For the models with $E_{51} = 10$, the following amount of $^{56}$Ni is found to be ejected: 
$\sim 2$ M$_{\odot}$, 1.5 M$_{\odot}$, and 1.2 M$_{\odot}$ for the 140 M$_{\odot}$, 
200 M$_{\odot}$, and 250 M$_{\odot}$ models, respectively. 
For these models, we estimate the luminosity powered by the $^{56}$Co decay following simple analytic models from the early \citep[e.g.,][]{arnett1982} to the late \citep[e.g.,][]{km2003} phases. 

We emphasize that our simplified treatment of the light-curve calculation allows only the first-order estimate, and the applicability to observational data requires several considerations. 
First, our assumption of the conversion efficiency of 30 per cent is calibrated for SNe IIn within a dense 
CSM \citep{Moriya14}. Generally, when the ambient density is low, the radiation does not reach to thermal emission \citep[e.g., ][]{chevalier2012}, and the efficiency as determined by the efficiency of non-thermal electron acceleration is as low as $\sim$ 1 per cent \citep{km2014}. 
Furthermore, in such a situation a bulk of the radiated energy is either in radio or X-rays, not in the optical wavelengths. Also, we neglect the radiation cooling feedback to hydrodynamics, motivated by previous studies where the adiabatic treatment provides a reasonable description of the shock wave dynamics for SNe IIn as calibrated by radiation-hydrodynamic simulations \citep{moriya2013,moriya2014}.

\subsection{Results}

\begin{figure}
\includegraphics[width=6cm,angle=270]{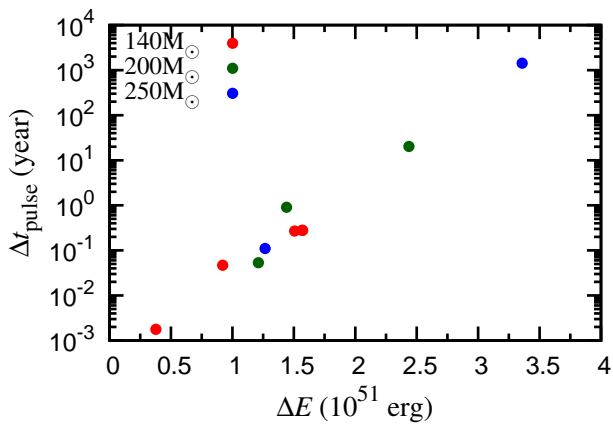}
\caption{The relation between the energy difference $\Delta E$ and the time interval of 
pulsations $\Delta t_{{\rm pulse}}$.
Red, green, and blue points indicate the 140, 200, and 250 M$_\odot$ models, respectively.
}
\end{figure}

Fig. 12 shows the calculated bolometric light curves. Fig. 13 shows the 
hydrodynamic evolution of the systems for the 140 M$_{\odot}$ model. 
The hydrodynamic evolution of the other progenitor models is qualitatively similar. 
Fig. 14 shows the position of the photosphere, as computed by the optical depth 
to electron scatterings in the shocked high temperature regions, 
for the 140 M$_{\odot}$ and 200 M$_{\odot}$ models. 
The system from the 250 M$_{\odot}$ model turns out to be optically thin except 
for the first year following the first pulse. 

Fig. 12 shows that the light curve reflects the successive ejections of the shells 
and the SN ejecta, and interactions between them. 
At the time of the SN explosion, the ambient density becomes high, 
at the level comparable to, or even higher than, the CSM created by a strong mass-loss 
with $\dot M \sim 0.1$ M$_{\odot}$ yr$^{-1}$ (see Fig. 13). 
This situation is similar to what is suggested for luminous SNe IIn and SLSNe reaching to 
$M_{\rm bol} \sim -20$ or even brighter \citep[e.g., ][]{Moriya14,moriya2014}. 
Indeed, we find that the calculated luminosities are similar to these events, or even brighter, 
reaching to $M_{\rm bol} \sim -21$ ($E_{51} = 1$) or $M_{\rm bol} \sim -23$ ($E_{51} = 10$) 
for the 140 M$_{\odot}$ model, $\sim -19$ ($E_{51} = 1$) or $\sim -21$ ($E_{51} = 10$) 
for the 200 M$_{\odot}$ model, and $\sim -18$ ($E_{51} = 1$) or $\sim -20$ ($E_{51} = 10$) 
for the 250 M$_{\odot}$ model. 

The difference in the shock-powered luminosity can be generally interpreted by the difference 
in the time-scale between the final shell ejection and the SN explosion, and also in the temporal 
history of the shell ejections. 
The 250 M$_{\odot}$ model has a relatively long gap between 
the final shell ejection and the SN explosion ($3.74$ yr), therefore the ambient density 
at the time of the SN explosion is much lower than the other cases. 
This leads to a lower luminosity in this model than the others. The shell ejection history is more 
complicated in the 140 M$_{\odot}$ model than the 200 M$_{\odot}$ model, 
and thus the deceleration of these pulses is more important in the former. 
Therefore, the higher density ambient material is created for the 140M$_{\odot}$ model, 
leading to the higher luminosity. 
Related to this, many `pre-SN precursors' are seen in the 140 M$_{\odot}$ model 
within a time-scale of a year, while the light curves of the other progenitor models are less 
complicated. 

The light curves are mainly powered by the interaction, but the contribution from the decay of 
$^{56}$Co (as initially produced as $^{56}$Ni at the SN explosion) can be significant. 
For the explosion energy of $E_{51} = 10$, a substantial amount of $^{56}$Ni 
($\sim$1--2 M$_{\odot}$) is created. 
In the 140 M$_{\odot}$ model, indeed the $^{56}$Co-decay power can exceed 
the interaction power, at a few months after the SN explosion and thereafter. 
Namely, we predict that the SN resulting from our scenario can change the properties and appearance according to the change in the power source, and in this case we predict 
that initially the SN should show signatures of strong interaction but later on it will look like SNe Ib/c. Note that the shocked region is optically thick even in the $^{56}$Co-power dominating phase, thus it may look like different from typical SNe Ib/c in having an optically thick, 
electron-scattering dominated atmosphere. 
In the 200 M$_{\odot}$ model, on the other hand, the $^{56}$Co-decay power can dominate initially 
before the interaction develops strongly. In this case, the $^{56}$Co-decay power may lead to a bright precursor having a character of typical SNe Ib/c, then a bright SN powered by the interaction follows.

\section{Discussion}

\subsection{Energy loss during PPI stage}

\begin{figure*}
%\epsscale{.50}
%\includegraphics[width=6cm,angle=270]{f3a.eps}
\includegraphics[width=5.5cm,angle=0]{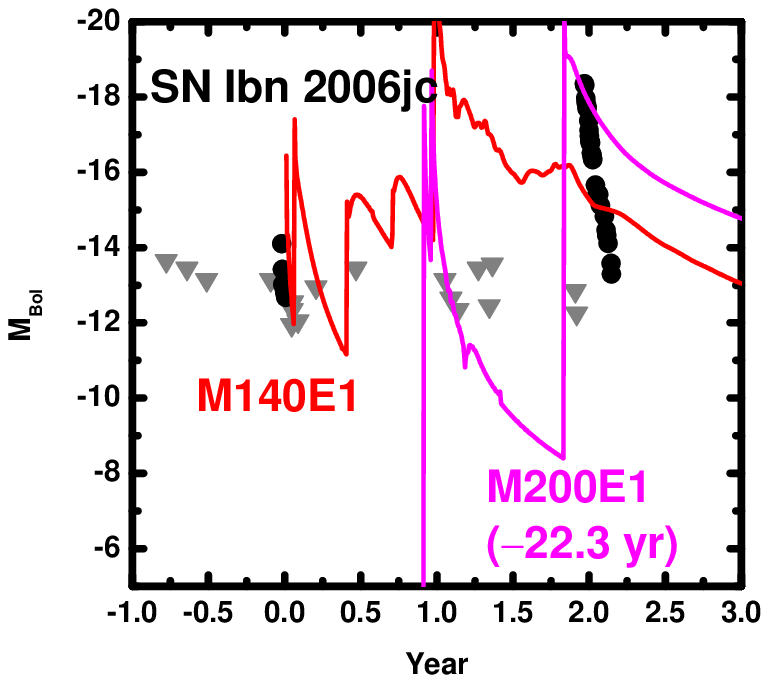}
\includegraphics[width=5.5cm,angle=0]{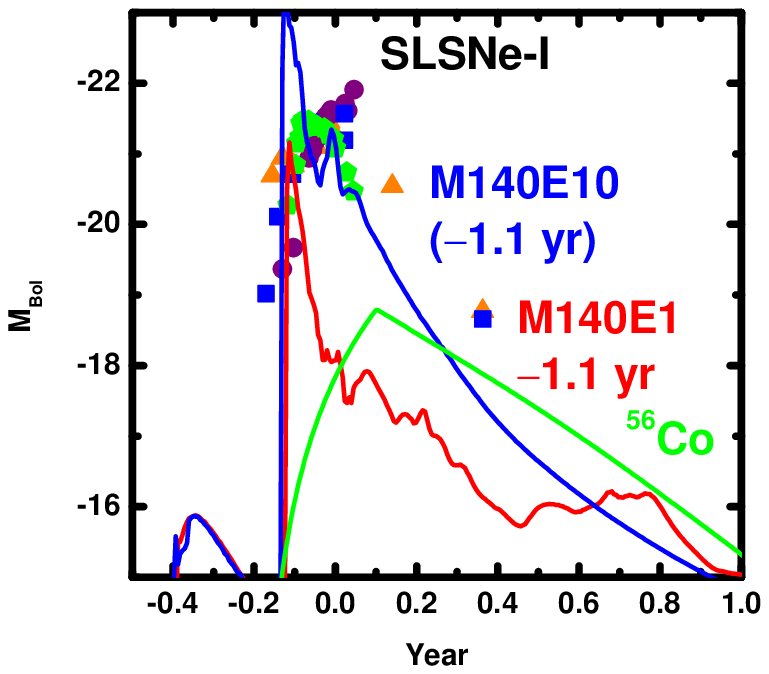}
\includegraphics[width=5.5cm,angle=0]{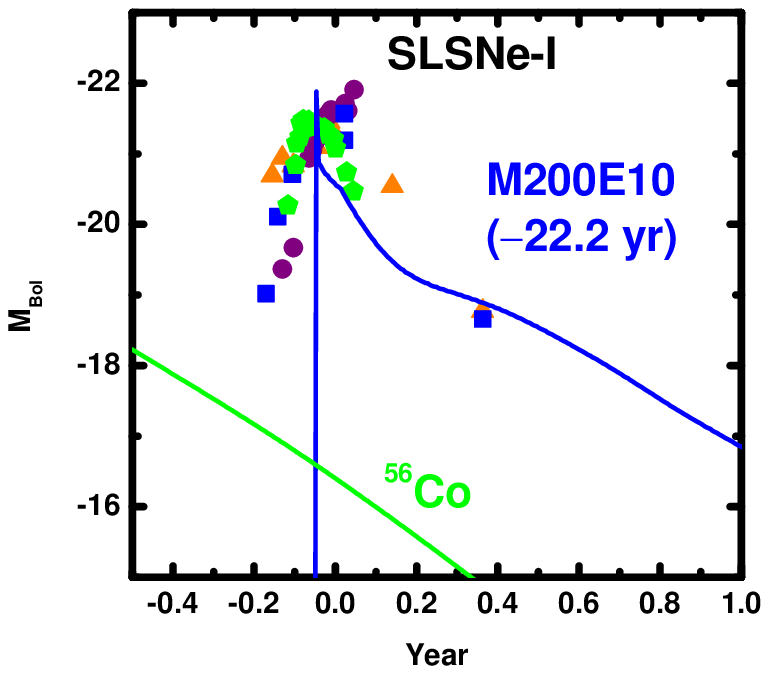}
\caption{Some comparisons between our model light curves with SNe: 
the {\it R}-band light curve of SN Ibn 2006jc \citep{Pastorello07} as compared to the bolometric 
light curves from models 140 M$_{\odot}$ ($E_{51} = 1$) and 200 M$_{\odot}$ 
($E_{51} = 1$) for left-hand panel,  
the {\it R}-band light curves of SLSNe-I 
\citep[orange-triangles: PTF09atu, purple-circles: PTF09cnd, blue-squares: PTF09cwl, 
green-pentagons: PTF10cwr, in][]{Quimby11}
as compared to the bolometric 
light curves for the 140 M$_{\odot}$ model ($E_{51} = 1$ and $10$) for center panel, 
and the 200 M$_{\odot}$ model (with $E_{51} = 10$) for right-hand panel. 
For the models with $E_{51} = 10$, the contribution of $^{56}$Co decay is shown by a green line. The models with $E_{51} = 1$ are shown by a red line while those with $E_{51} = 10$ are shown by a blue line, except for the 200 M$_{\odot}$ model in the left-hand panel (shown by a magenta line).  
}
\end{figure*}

We discuss time intervals and strength of pulsations.
We find from Figs. 3, 8, and 10 that 
if the minimal temperature in each pulse is lower, the time interval is longer.
In each pulse, a star is most loosely bound when the central temperature
has a minimal value.
We define the total energy at the minimal central temperature to be $E(T_{{\rm min}})$.
Then, the star gradually contracts and, at the same time, the energy is carried away 
by neutrinos.
When the star dynamically contracts and just before rapid nuclear burning starts, 
the total energy of the star should have a minimal value $E_{{\rm min}}$.
We investigate the relation of the energy difference 
$\Delta E = E(T_{{\rm min}}) - E_{{\rm min}}$ to the time interval of pulsations 
$\Delta t_{{\rm pulse}}$.
Fig. 15 shows the relation between the energy difference $\Delta E$ and the time interval
$\Delta t_{{\rm pulse}}$.
Although the stellar-mass dependence is unclear, the time interval  strongly depends on 
$\Delta E$.
Strong temperature dependence of the neutrino-cooling rate brings about the shown dependence.
We also see a rough correlation between $\Delta E$ and the ejected mass.
The energy difference $\Delta E$ would be an indicator of the strength of pulsations.

In our calculations, the first pulsation seems to be weak for all models.
In the 200 and 250 M$_\odot$ models, the second pulsation is the strongest.
This trend is somewhat different from the results in the supplementary information of 
\citet{Woosley07}.
They showed that the He star models with $M_{{\rm He}} \ge 54$ M$_\odot$ indicated the
longest time interval in the first pulsation.
Typical kinetic energy of ejecta in our models seems to be smaller than that of them.
We have not found a reason of the difference.
The kinetic energy may relate to the time evolution of the density structure during
pulsations complicatedly.
We do not see a clear relation between the kinetic energy of ejecta and the pulse interval.
Our models do not have the He-rich envelope.
The CO-rich envelope may be bound stronger than the envelope of He-star models.

\subsection{Implications for luminous SNe}

For an illustration purpose, we compare our bolometric light curves with the $R$-band light curves of 
some SNe in Fig. 16. 
Since our model is essentially hydrogen-free, we restrict the comparison to a class of 
SNe Ib/c for which the hydrodynamic interaction has been suggested to be a source of the 
luminosity (among several other interpretations). 
The comparison includes the following SNe: (1) SN Ibn 2006jc as a prototypical SN showing the 
strong interaction between the SN ejecta and H-free environment 
\citep{Pastorello07,anupama2009} \citep[but see][for different interpretation]{tominaga2008}, 
and (2) SLSNe-I, which are characterized by a huge luminosity 
exceeding $M_{\rm R} \sim -21$ and by H-free spectra \citep{Quimby11, Gal-Yam12}, 
for which not only the progenitor system but also the emission process have not been clarified yet 
\citep[e.g., a magnetar-powered model][]{kasen2010} \citep[see also][]{km2007}. 
Note that SLSNe-I do not show obvious spectrum signatures of interaction, but how the interaction 
between the SN ejecta and H/He-poor CSM manifests itself in spectra has not been clarified:
it may not necessarily show obvious strong emission lines \citep{Sorokina15}.

The models do not well reproduce the light curve of SN 2006jc. The time-scale of 2 yr between the precursor and the SN explosion is roughly right as the time-scale expected for the pulsational activities in the 140 M$_{\odot}$ or 200 M$_{\odot}$ progenitor. Also, 
the peak luminosity is roughly consistent, for $E_{51} = 1$, to the observed one, which means that 
the PPISN model can provide the environment similar to that around SN 2006jc without fine-tuning. 
However, the model light curve after the SN decays much slower than the observations, which is also 
another indication of properties of the environment. There is thus a difficulty to explain the relatively 
rapid decay as seen in SN 2006jc by the PPISN model. 

As for SLSNe-I, the explosion 
energy as large as $E_{51} = 10$ is required to explain the luminosity and duration of these SNe, 
as expected. 
While details are not fitted by our models, these models provide qualitatively 
good similarity to the observed light curves. 
For both 140 M$_{\odot}$ and 200 M$_{\odot}$ models, the luminosity and decay time-scale 
are explained reasonably well, without any fine-tuning. Another interesting property is seen in the 
140 M$_{\odot}$ model. 
In this model, the $^{56}$Co-decay dominates the power input from $\sim 3$ months after the 
explosion, and then the SN should look like a SN Ib/c. Indeed, at least some SLSNe-I with 
such late-time observations available, they show spectra similar to SNe Ib/c\citep{Quimby11,nicholl14,chen15}. 
We also note that while the CSM interaction is disfavoured as a model for some SLSNe-I based on a rapidly decaying model light curve as compared to observations \citep{nicholl14}, the PPISN model could evolve more slowly than a fiducial interaction model either owing to the contribution by the $^{56}$Co power (140 M$_{\odot}$; see centre panel of Fig. 16) or owing to a complicated pre-SN CSM structure (200 M$_{\odot}$; see right-hand panel of Fig. 16). 

The 140 M$_{\odot}$ model also predicts the pre-SN activities at the level of $M_{\rm bol} \sim -16$, 
which happen several times in 1 yr time-scale. \citet{leloudas2012} found 
a probable precursor in SLSN-I 2006oz, at 6--10 d before the beginning of the main part 
of the light curve, or $\sim$30--40 d before the peak luminosity. 
The time-scale may be consistent with our model, 
while the luminosity of the precursor ($\sim -19$ mag) far exceeds our model prediction 
($\sim -16$ mag). 
Details will be sensitive to the progenitor mass, and further investigation for different mass 
is required. 
In any case, within our scenario we generally expect a larger number of multiple pre-SN outbursts 
for a brighter SN (linked by the density of the environment), and SLSNe-I following this path, 
if they exist, should show such activities. 
Therefore, future observations to catch pre-SN outbursts are important to discriminate our models from various proposed models for SLSNe-I. For SLSNe-I at the redshift 
of $z = 0.2$, the pre-SN outburst of the 140 M$_{\odot}$ model should reach to an apparent magnitude of $m_{\rm bol} \sim 24$, thus this can be investigated by surveys with 4m-class telescopes. Typical sensitivities by surveys with 8m-class telescopes (e.g., Subaru/HSC, LSST) will be sufficient for the detection of such outbursts up to $z \sim$0.5--0.6. For example, the ongoing 
Subaru/HSC-deep survey is expected to detect $\sim 10$ SLSNe below $z \sim 1$ 
\citep{tanaka2012}, and it is designed to visit the same field twice a month for continuous 3 
months. With this cadence we could detect at least the outburst following the ejection of the final shell before the SN explosion (Figs. 12 and 16).

If a very massive star collapses and does not explode after the PPI stage, 
only precursor events before the collapse could be observed.
In the 140 M$_\odot$ model, multiple peaks with $M_{\rm bol} \sim -16$ could be
observed in the light curve (see the left-hand panel of Fig. 12).
This feature may be similar to the precursor of SN 2006jc.
Two luminous events with $M_{\rm bol} \sim -16$ and $-18$ will be observed
in the interval of $\sim 20$ yr for the 20 M$_\odot$ model.
The 250 M$_\odot$ model will give a luminous event achieving $M_{\rm bol} \sim -20$.
If a strong pulsation occurs like the 200 and 250 M$_\odot$ models, the luminosity of the
luminous event by the interaction can be luminous and could achieve that of an SLSN.

Recently, an SLSN-I showing H$\alpha$ emissions in its late-time spectra
has been observed.
The SN, iPTF13ehe, was observed as an SLSN-I and its light-curve has typical SLSN-R features
\citep{Yan15}.
They considered that the broad H$\alpha$ emission line that appeared in late phases is formed 
by the interaction between SN ejecta and the $< 30$ M$_\odot$ H-rich CSM shell, 
which has been ejected by the PPI.
They also modelled the nebular spectra, which suggest 
$M$($^{56}$Ni)$ \sim 2.5$ M$_\odot$.
Such a large amount of $^{56}$Ni production accompanying the PPI is possible only 
by an energetic SN explosion induced by the CC of an Fe core in a massive CO core.
This observational feature could suggest that substantial $^{56}$Ni ejection is indeed 
possible by an energetic CCSN from a massive CO core 
as discussed in \citet{Umeda08}, \citet{Yoshida14} and this study.

A possibility of strong X-ray emissions from the interaction between 
the SN forward shock and dense CSM (due to pulsational mass-loss) has been investigated 
by \citet{Pan13}.
In our 140 and 200 M$_\odot$ models, the strong interaction takes place at $R \sim 10^{16}$ cm,
and thus the interacting region is optically thick, forming a photosphere (see Fig. 14).
This is in accordance with the estimate by \citet{Pan13}.
In this case, most of the emitted X-rays will be converted into optical emission within the interaction region.
Still, the region immediately behind the forward shock can be optically thin avoiding thermalization, and
thus a fraction of the luminosity may be emitted in X-rays.
This will be negligible as compared to the optical luminosity, but could still be considered as a strong
X-ray emitter given by the large bolometric luminosity.
For the 250 M$_\odot$ model, indeed the interacting region is optically thin to electron scatterings
for most of the time.
Therefore, it is possible that a large fraction of the luminosity would be emitted as X-rays.
However, the ejecta will still be optically thick (which we did not take into account in our simulations),
and will convert about a half of the X-ray to optical emission.
Therefore, in this case the SN could be a very strong X-ray emitter, but still will not change our result
about the optical luminosity significantly.
In any case, it is possible that the PPI SNe could be luminous in X-rays and possibly be observed
in the high energy emission.

\section{Summary}

We investigated the evolution of the 140, 200, and 250 M$_\odot$ models with $Z=0.004$, 
which had become 54.1, 58.7, and 61.0 M$_\odot$ after the C-burning,
including mass ejection induced by PPI.
Then, we calculated SN explosions with two cases of explosion energy.
We also calculated light curves created by the interaction of the CSMs ejected by the PPI
and that between the CSM and the SN ejecta.
The obtained light curves were applied to optically luminous transients.
We summarize the main results as follows.

The 140, 200, and 250 M$_\odot$ models experienced six, four, and three pulsations
during the PPI stage.
A part of the outer envelope exceeded the escape velocity and, thus, ejected from the stars.
The final stellar masses of these models are 50.10, 53.35, and 53.16 M$_\odot$.

The surface composition changes by the mass ejection during the PPI stage.
The final He mass fraction at the surface is $\sim 2 \times 10^{-3}$ for the 140 M$_\odot$ model
and much smaller for the other two models.
If these stars explode, they will be observed as SNe Ic.

Larger CO-core model experiences the PPI stage with fewer pulsation numbers and longer period.
This trend is consistent with the result in \citet{Woosley07}.
The PPI period is almost determined by the strongest pulsation in the 200 and 250 
M$_\odot$ models.

The interactions of the PPI ejecta with the CSM formed by previous ejections
make optically luminous transients.
The 140 M$_\odot$ model reaches $M_{{\rm bol}} \sim -16$ to $-18$ for several times.
The 200 and 250 M$_\odot$ models become bright up to $M_{{\rm bol}} \sim -18$ and $-20$,
respectively, even without an energetic SN.
If an SN follows the CC, then the luminosities produced by the interactions of the CSM 
and the SN ejecta are higher than those of the CSM interactions.
The obtained luminosities become comparable to SLSNe-I.

Although there are still difficulties such as reproducing short SN decay time, the CSM
interaction during the PPI stage and the following SN could be tested by light curves of SLSNe-I
in future surveys.

\section*{Acknowledgements}

We thank Tony Pan for useful comments on X-ray emission from PPI SNe.
This work has been partly support by the grants-in-aid for Scientific Research
(24244028, 26400271, 26800100) from the MEXT of Japan
and WPI Initiative, MEXT, Japan.

\bsp

\label{lastpage}

\end{document}